# Ballistic vs. Diffusive Transport in Current-Induced Magnetization Switching

N. Theodoropoulou, A. Sharma, W.P. Pratt Jr., and J. Bass

*Department of Physics and Astronomy, Michigan State University, East Lansing, MI 48824-2320, USA*

M. D. Stiles,

*Center for Nanoscale Science and Technology, National Institute of Standards and Technology, Gaithersburg, MD 20899-6202, USA.*

Jiang Xiao*

*School of Physics, Georgia Institute of Technology, Atlanta, Georgia 30332-0430, USA*

We test whether current-induced magnetization switching due to spin-transfer-torque in ferromagnetic/non-magnetic/ferromagnetic (F/N/F) trilayers changes significantly when scattering within the N-metal layers is changed from ballistic to diffusive. Here ballistic corresponds to a ratio $r = \lambda/t \geq 3$ for a Cu spacer layer, and diffusive to $r \leq 0.4$ for a CuGe alloy spacer layer, where $\lambda$ is the mean-free-path in the N-layer of fixed thickness $t = 10$ nm. The average switching currents for the alloy spacer layer are only modestly larger than those for Cu. The best available model predicts a much greater sensitivity of the switching currents to diffuse scattering in the spacer layer than we see.

Current-Induced Magnetization Switching (hereafter just current-induced switching) due to spin-transfer-torque in ferromagnetic/non-magnetic/ferromagnetic (F1/N/F2) trilayers was predicted in 1996 [1,2], and has been studied experimentally since 1998 [3-7]. However, published models have not yet been shown to quantitatively reproduce measured switching currents [8-16]. It has been difficult to critically evaluate these models, because comparing calculations with data for samples having several different components is usually complicated. In the present experiments, we simplify by using a test based on a ratio of results from measurements on samples in which only the N-layer is modified, with the main effect being just to change the scattering in this layer from ballistic to diffusive. Specifically, we compare the prediction of a Boltzmann equation model [16] to the measured ratio of switching currents ($X = \Delta I(CuGe)/\Delta I(Cu)$), for sample sets that are nominally identical in every way (including the leads), except for changing a fixed 10 nm thickness of the N-metal from nominally 'pure' Cu to a dilute Cu(5 at.% Ge)—hereafter just CuGe—alloy. Here $\Delta I = I^+ - I^-$ is the difference between the positive ($I^+$) and negative ($I^-$) switching currents shown in the upper left panel of Fig. 1, labelled H = 0. This Boltzmann model has the advantages of interpolating between the ballistic and diffusive limits, and generally including other approaches [8-15] as limiting cases. We chose CuGe because Ge in Cu: (a) adds a large resistivity per atomic percent Ge ($\approx 3.7$ μΩcm/at.% [17]), thereby greatly shortening the mean-free-path, λ; (b) has a small spin-orbit cross-section in Cu ($\sigma_{so} \approx 5.2 \times 10^{-19}$ cm$^2$ [18]), giving weak spin-flipping and thus a long spin-diffusion length, $\ell_{sf}^N$; and (c) is soluble in Cu to $\approx 10$ % [19].

As shown in Table I, with Cu, λ is at least 3 times the layer thickness of 10 nm (ballistic transport). With CuGe, λ is less than 40% of 10 nm (diffusive transport). We can thus determine, at both 295K and 4.2K whether: (1) just changing the scattering in the N-layer from ballistic to diffusive significantly affects the switching current, $\Delta I$; and (2) how well the Boltzmann equation model of current-induced switching describes the ratio of switching currents. By using a ratio, we expect that most, if not all, of any systematic errors will cancel.

Our sputtered Cu(80)/Py(24)/N(10)/Py(6)/Au(10)/Cu(20) ≈ 70 nm x 130 nm nanopillars were prepared by a combination of optical and e-beam lithography plus subtractive ion milling as described in refs. [20,21]. Layer thicknesses are in nm, and Py = $Ni_{1-x}Fe_x$ with x ~ 0.2. To minimize magnetic coupling between the two Py layers, so we can simply measure switching at H = 0, we left about half of the middle N-layer (Cu or CuGe) and all of the bottom Py and Cu layers unpatterned. With fully patterned nanopillars, studying switching would require applying a field to offset the dipolar coupling field, which is not uniform across the sample. Ion-beam etching is done in a chamber with background pressure ≈ $10^{-5}$ Pa, and we deposit our insulating SiO without breaking vacuum.

Strictly, the calculations in refs. [8-16] give the torque, which is most closely related to the onset of dynamical instability. Since this onset is hard to determine, we measure the currents at which sharp switching occurs (illustrated at both 295K and 4.2K for one sample each of CuGe and Cu in Fig. 1), neglecting any 'pre-switching' structure in the hysteresis curves. This process minimizes ambiguity, and is the same for samples with Cu and CuGe. At 295 K, the great majority of our switchings were sharp and complete; at this temperature we include in our discussion data for a total of 36 samples with Cu and 29 for CuGe (see Table II below). At 4.2 K, sharp switching was less prevalent; there we include data (**bold** in Table II) for only 7 samples of Cu and 6 of CuGe.

Crucial to any quantitative comparison of data and theory are checks for possible systematic errors in the data. To make such checks, we include in Table II results for the 13 Cu and CuGe samples of main interest at both 295K and 4.2K, and the other 52 samples with Cu and CuGe at 295 K. We discuss the comparisons of these results below.

One of these checks for systematic errors involves results from our previous study at 295 K on samples using F1 = F2 = Py and N = Cu, or Cu alloyed with 5 at. % Pt sandwiched between Cu layers, to test effects of spin-flipping within the N-layer [20]. Pt produces strong spin-flipping in Cu [18] with only modest elastic scattering [17]. For those samples, the inverse critical switching currents, $(1/I^+)$ or $(1/I)$, were found to be directly proportional to the change in resistance $\Delta R = R(AP) - R(P)$ between the states with moments of the two F-layers anti-parallel (AP) or parallel (P) to each other [20]. As linear fits to the data of ref. [20] were compatible with zero intercept for both positive and negative (1/$I$), we recast the data into the form $\Delta R \Delta I$ = constant, where $\Delta I = I^+ - I^-$ was defined above (because of uncertainties in the data, we do not claim this recasting as 'definitive'—there might be small non-zero intercepts). We used this relationship in two ways: (a) to check if our new values of $\Delta R \Delta I$ with Cu agree with earlier ones [20-22]—they do; and (b) to check if the values of $\Delta R \Delta I$ for CuGe agree with those for Cu—Table II shows that they do. These agreements at both 295 K and 4.2 K give us confidence that any systematic errors in our data should not be large.

To specify the differences between our samples with Cu and CuGe, and to calculate $X$, we need estimates of the mean-free-paths, $\lambda$, and spin-diffusion lengths, $\ell_{sf}^N$, for our samples. To estimate $\lambda$, we use the free-electron relation between resistivity, $\rho$, and $\lambda$, in Cu: $\lambda\rho$ = 660 n$\Omega$m$^2$ [17,23] along with values of $\rho$ measured separately on 200 nm thick films using the van der Pauw technique [24]. Values of $\rho$, $\lambda$, and the ratio $r = \lambda/t$ are given in Table I. To correct for spin-flipping, we estimate $\ell_{sf}^N$ from the Monod and Schultz collection of cross sections $\sigma_{so}$ measured by electron spin resonance [18]. For Ge in Cu, they list $\sigma$ = 5.2 x 10$^{-19}$ cm$^2$, which yields a spin-flip length for our CuGe--alloy of $\lambda_{sf} \approx$ 4500 nm. We estimate the 4.2K spin-diffusion length for our CuGe--alloy from the Valet-Fert Eqn. $\ell_{sf}^N = \sqrt{(\lambda \lambda_{sf})/6}$ = $\sqrt{[(4.1\pm 0.2)x(4500\pm 450)/6}$ $\cong$ 55$\pm$5 nm [25].

From data such as those in Fig. 1, we derived values at 295 K and 4.2 K of $R$(AP), $\Delta I$, $\Delta R$, and the product $\Delta R \Delta I$ for nanopillars with both Cu and CuGe. The average values for these quantities are collected together in Table II. In each case, the number of samples measured is given in parentheses. Values of $R$(AP) are for complete samples, including the wide Cu and Py layers at the bottom and the wide Au and Cu layers at the top. Compared to Cu, the larger resistivity of CuGe should increase $R$(AP) by ~ 0.1 $\Omega$ at both 295K and 4.2K, well within our uncertainties.

From the values of $\Delta I$ in Table II, we derive the ratios $X$. At 295 K we find $X = 1.4 \pm 0.4$ for the 13 samples and $X = 1.4 \pm 0.2$ for the other 52 samples. At 4.2 K we find $X = 1.1 \pm 0.3$ for the 13 samples. We take the agreement on $X = 1.4$ at 295 K, as well as those of $\Delta R \Delta I$ in Table II already noted, as evidence that our 13 selected samples are representative of both our larger set of 52 samples and our prior data on Cu [20-22].

We calculate $X$ for a Boltzmann equation model [16], in which the torque (per unit current) is determined as a function of angle by solving the Boltzmann equation numerically in a spin-valve, using independently determined parameters for the nanopillar components [26]. The model includes both spin-conserving and spin-flip scattering in all layers, and smoothly changes from a diffusive to a ballistic regime as the mean free path of the spacer layer changes. The critical current is inversely proportional to the slope of the torque curve, giving $\Delta I \propto |d\tau(P)/d\theta|^{-1} + |d\tau(AP)/d\theta|^{-1}$. The values of $\lambda_N$ and $\ell_{sf}^N$ for Cu and CuGe listed above, together with the standard parameters [26], give $X = 2.0 \pm 0.13$, due mainly to the much shorter mean free path (much larger $\rho_N$) for CuGe. The uncertainty, most sensitive to the uncertainties in the thickness of the CuGe layer, is computed from the experimental uncertainty in the input parameters and the computed sensitivity of the result to each parameter. As a cross-check, we also calculated $X$ using a closely related model by Fert et al. [14] with the same input parameters, and found a very similar result, $X = 2.2$.

Our calculated value of $X = 2.0$ is considerably larger than our experimental values of $X$ at both 295 K and 4.2 K. We can identify several possible reasons for this discrepancy, although we do not find any compelling.

One set of complications is due to thermal effects. Both the transport and magnetic dynamics are calculated at zero temperature. We expect any temperature dependence of torque to be weak, because the only variations with temperature in electronic properties should be changes in scattering rates. We also expect the dominant thermal effects to come from thermally activated dynamics, so that, absent complications, the models should be compared with our 4.2 K data.

However, at 4.2 K, magnetic anisotropies in the F-layers might lead to more complicated reversal processes where $\Delta I$ is not as closely related to the torque as assumed. Also, at 4.2 K, adventitious antiferromagnetic NiO on the surfaces of Py-layers in nanopillars was reported to affect switching [27], by producing local exchange-biasing that generated both large variations in the switching field $H_s$, and occasional multiple steps. While we do see variations in $H_s$ in both Cu and CuGe sample sets, they occur at both 4.2 K and 295 K. Since 295 K is well above the Neel temperature of thin NiO layers, variations there cannot be attributed to pinning. We also occasionally see steps at 4.2K. But we omit any such samples from our analysis. Lastly, as any effects of NiO should be similar for Cu and CuGe, we doubt that their presence would strongly bias our measured ratios. Fortunately, the issue of temperature is not crucial to our conclusions, as our 4.2K and 295K values of $X$ overlap, with a 'common' value of $X = 1.3$, to within mutual uncertainties.

One possible cause of the discrepancy between the model and the measurements is that we measure reversal, whereas the model calculates the initial dynamical instability. In addition, the model assumes a macrospin, and the experimental reversal mode may be non-uniform [28]. However, we do not expect these complications to differ much for the Cu or CuGe spacers.

Another issue is the geometry. The calculation assumes that the devices are wide wires with uniform current flow. As noted above, so that we could measure switching at H = 0 (and have it insensitive to H around H = 0), the actual device structure is more complex, leading to non-uniform current flow and more complicated behaviors of spin accumulation and spin current. Berger [29] and Hamrle et al. [30] have discussed corrections for such non-uniformity, which conceivably might differ for Cu and CuGe. However, since the cross sectional dimensions of the nanopillars are larger than the layer thicknesses and the mean-free-paths (except for Cu at 4.2K), we don't expect a 3-dimensional treatment to give very different changes for the two sets of samples.

Finally, there could be other errors in the theoretical approach. The model approximates all Fermi surfaces as spheres with the same radii, idealizes the interfaces as specular, and ignores spin-flip scattering at the interfaces. We don't yet know how much change could be produced by relaxing any of these approximations. Since the nanopillar resistances are comparable to those of the leads, the resistances and magnetoresistances can't be used to constrain the calculations or independently check the transport parameters, which might not be appropriate for nanopillars. Some of these issues might be addressed in quantum mechanical calculations [31,32], where, however, it is hard to include diffuse scattering.

To summarize, to see if spin-torque differs for ballistic vs diffusive scattering in the N-metal, and to test the best available model of spin-transfer-torque produced Current-Induced Magnetization Switching, we measured the simple ratio $X = \Delta I(\text{CuGe})/\Delta I(\text{Cu})$ at 295 K and 4.2 K for samples with Cu giving ballistic transport (i.e. ratio $r = \lambda/10 \text{ nm} \geq 3$) and samples with CuGe giving diffusive transport ($r \leq 0.4$), with no spin-flipping as electrons transit the Cu, and modest spin-flipping for CuGe. At 295K we find $X = 1.4 \pm 0.2$ and at 4.2K, $X = 1.1 \pm 0.3$. These values are not far from unity, indicating that there isn't much difference in spin-torque for ballistic or diffusive scattering in the N-metal. However, they are noticeably smaller than expected from a Boltzmann equation [16] model, or the related Fert et al. one [14]. While experimental issues cannot be ruled out, we think it more likely that the calculations fail to capture some essential feature of the experiment. The calculations could be improved by adding micromagnetics, more realistic treatment of transport in the sample geometry, and first principles transport calculations including diffuse scattering. But each of these advances is computationally intensive and it is very hard to do any two with present computers. It is also not clear that any will give different results for Cu and CuGe spacers. We, hope that our results will provide a benchmark for theoretical analyses to advance understanding of spin transfer torques.

We gratefully acknowledge support from the MSU Keck Microfabrication Facility and the NSF via grant DMR-05-01013.

* Present Address: Kavli Institute of NanoScience, Delft University of Technology, 2628 CJ, Delft, The Netherlands.

Table I. Resistivities, $\rho$, mean-free-paths $\lambda$, ratios $r = \lambda/10$ nm and $r^* = \ell_{sf}^N/10$ nm for N = Cu and CuGe.

| Metal | T(K) | $\rho(\mu\Omega cm)$ | $\lambda$ (nm) | r | r* |
|---|---|---|---|---|---|
| Cu | 293 | 2.1±0.1 | 31±2 | 3.1 | ~29 |
| CuGe | 293 | 18.0±0.9 | 3.7±0.2 | 0.37 | ~5 |
| Cu | 4.2 | 0.5±0.1 | 132±20 | 13.2 | ~100 |
| CuGe | 4.2 | 16.2±0.7 | 4.1±0.2 | 0.41 | ~5 |

Table II: $R$(AP), $\Delta I$, $\Delta R$, and $\Delta R\Delta I$ for Cu and CuGe at 295 K and 4.2 K. Numbers in parentheses in the first column indicate the number of samples in each data set. Uncertainties are two standard deviations of the mean. Bold indicates the 13 samples with sharp switching at 4.2 K, and standard font indicates the other 52 samples. The agreements for different sample sets within mutual uncertainties suggests that systematic errors in $R$(AP), $\Delta R\Delta I$, and especially $\Delta I$, for the 13 new samples are not large.

|  | $R$(AP)($\Omega$) | $\Delta I$(mA) | $\Delta R$(m$\Omega$) | $\Delta R\Delta I$(m$\Omega$A) |
|---|---|---|---|---|
| **295 K** |  |  |  |  |
| **Cu (7)** | **2.2±0.8** | **5.5±1.0** | **56±16** | **0.31±0.11** |
| **CuGe (6)** | **2.2±1.0** | **7.5±1.6** | **41±4** | **0.31±0.07** |
| Cu (29) | 2.0±0.4 | 6.0±0.8 | 57±14 | 0.34±0.10 |
| CuGe (23) | 1.5±0.2 | 8.3±0.8 | 35±6 | 0.29±0.06 |
| **4.2 K** |  |  |  |  |
| **Cu(7)** | **1.9±0.8** | **9.4±1.4** | **114±30** | **1.07±0.32** |
| **CuGe(6)** | **1.9±1.0** | **10.2±2.0** | **87±14** | **0.89±0.22** |

**Figure Captions.**

Fig. 1. *R*(*H*) and *R*(*I*) at 295K (top) and 4.2K (bottom) for one of the six selected CuGe nanopillars (left) and one of the seven selected Cu nanopillars (right). Arrows in the upper left indicate Δ*R* and Δ*I*.

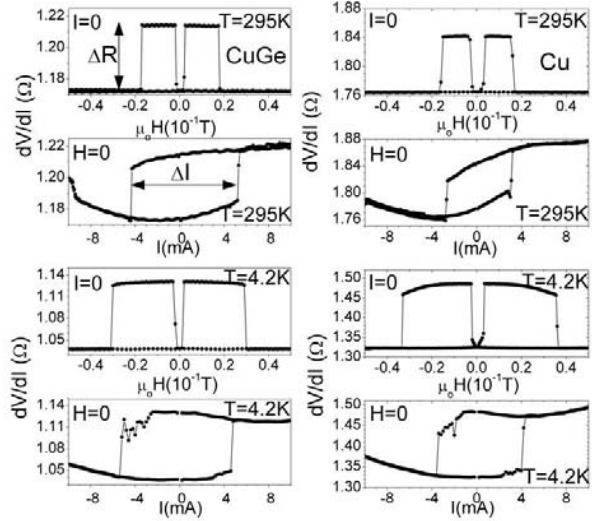

Figure 1